\documentclass[12pt]{iopart}
\usepackage{iopams}  

\usepackage{graphicx}
\usepackage{dcolumn}
\usepackage{bm}

\usepackage{color}
\usepackage{multirow}

\usepackage[FIGTOPCAP,raggedright,normalsize,tight]{subfigure}

\begin{document}

\title[Mechanical properties of nanosheets and nanotubes ...]{Mechanical properties of nanosheets and nanotubes using a new geometry independent volume definition}

\author{Philipp Wagner$^1$, Viktoria V. Ivanovskaya$^1$, Mark J. Rayson$^2$, Patrick R. Briddon$^3$, Christopher P. Ewels$^1$}
 \address{$^1$Institut des Mat\'eriaux Jean Rouxel (IMN), Universit\'e de Nantes, CNRS UMR 6502, 44322 Nantes, France}
 \address{$^2$Dept. Eng. Sciences and Mathematics, Lule\r{a} University of Technology, 97187 Lule\r{a}, Sweden}
 \address{$^3$School of Electrical and Electronic Engineering, University of Newcastle, Newcastle upon Tyne, NE 1 7RU, United Kingdom}
 \ead{\mailto{philipp.wagner@cnrs-imn.fr}, \mailto{chris.ewels@cnrs-imn.fr}}

\date{\today}

\begin{abstract} 
Cross-sectional area and volume become difficult to define as material dimensions approach the atomic scale.  This limits transferability of macroscopic concepts such as Young's modulus. We propose a new volume definition where the enclosed nanosheet or nanotube average electron density matches that of the parent layered bulk material. We calculate Young's moduli for various nanosheets (including graphene, BN and MoS$_2$) and nanotubes. Further implications of this new volume definition such as Fermi level dependent Young's modulus and out-of-plane Poisson's ratio are shown. 

\end{abstract}

See also \JPCM 25, (15) 155302, 2013 

\maketitle

\section{Introduction}

While mechanical reinforcement with single-wall carbon nanotubes (SWCNTs) has been a hot topic since the 1990s \cite{Baughman2002}, recently interest is also growing in individual-layer or few-layer based nanomaterials such as graphene, BN and MoS$_2$ \cite{Stankovich2006,Coleman2011short,Neto2011,Wagner2011a}.  Bulk mechanical properties are commonly specified using well defined parameters such as Young's modulus $E$ (see also \ref{Formule_E} and \cite{submat}).
When making the transition to nanoobjects this leads to complications, since the object boundaries and hence volume and cross-section have no general and transferable definition.
Thus while elastic tensors remain unambiguously defined at these scales, the conversion of both experimental and theoretical strains and forces into mechanical constants such as Young's modulus require a definition of mechanically active volume.\\
To date no such generalised and transferable volume definition exists.  A common approach is to use geometric ``macroscopic" volume models such as a rectangular slab for flat graphene or an empty cylinder for SWCNTs. However literature values chosen for the thickness $t$ of the graphene slab or SWCNT cylinder range from $t=0.6-3.4$ \AA \; \cite{Huang2006,Scarpa2009}, leading to wildly different volumes or cross-sections. The result is a wide scatter in reported values of the in-plane Young's modulus for graphene and the axial Young's modulus for SWCNTs, between $0.5-5.0$ TPa \cite{Huang2006,Scarpa2009}.
Currently the most common approach for graphene is to consider it as a uniform slab with thickness of the interlayer spacing of graphite (3.35~\AA). When both theory and experiment adopt this same value, the result is reasonably matching values of the in-plane Young's modulus between theory $0.86$ \cite{Pei2010} - $1.11$ TPa \cite{VanLier2000} and experiment $1.0$ \cite{Lee2008} - $1.02$ TPa \cite{Lee2009}.  
Simply transferring the graphite inter-layer distance to the cylinder thickness for SWCNTs provokes questions about the influence of curvature on the volume \cite{Akdim2003,Hernandez1998}, especially for narrow nanotubes. To date all such geometric approaches have in common the lack of a conceptual framework required for its generalisation to other related structures.\\  
Volume can alternatively be defined based on a sum of spherical overlapping atomic radii, such as covalent or Van der Waals radii \cite{Pauling1947,Pyykkoe2009,Bondi1964,Mantina2009}.  However by drawing on a library of pre-existing small molecules rather than considering the precise system in hand, such definitions once again suffer from a lack in transferability.  Notably $\pi$-bond systems are very poorly represented via Van der Waals radii \cite{Warburton2008}. 
Thus to date there is no general method to describe mechanically active nanoobject volume, capable of describing different kinds of structures without introducing various empirical or experimental parameters.\\ 
In this article we present a new geometry independent, parameter free and transferable volume definition based on the electron density distribution in the material, accessible from general density functional (DFT) calculations. We apply this to calculations of Young's modulus and Poisson's ratio of nanosheets and single-wall carbon nanotubes. This new definition provides a robust, reliable, quantitative basis for future mechanical studies of nanomaterials.\\

\section{Method}

In the following study we use DFT calculations under the local density approximation, implemented in the \textit{AIMPRO} code \cite{Rayson2008,Rayson2009,Briddon2011}. Relativistic pseudopotentials are included via the Hartwigsen-Goedecker-Hutter scheme \cite{Hartwigsen1998}. The basis consists of Gaussian function sets multiplied by polynomial functions including all angular momenta up to maxima $p$ ($l=0,1$) and $d$ ($l=0,1,2$) \cite{Goss2007}. For example, for carbon a $pdddp$ basis set was used, resulting in $38$ independent functions. Periodic boundary conditions are used, with system-dependent plane wave energy cutoffs up to $175$ Ha (Ha: Hartree energy), and a non-zero electron temperature of $kT=0.04$ eV to create electronic level occupation. The $k$-point grids were sufficiently fine to give energies converged to better than $10^{-7}$ Ha. Atomic positions and lattice parameters were geometrically optimised until the maximum atomic position change in a given iteration dropped below $10^{-6}$ a$_0$ (a$_0$: Bohr radius). To avoid interaction, supercell sizes were chosen such that the distance between structures was larger than $22.7$ a$_0$ ($12$ \AA). For Young's modulus calculations we apply small strains $\epsilon$ ($\pm 0.5,\pm 1.0, \pm2.0$ \%) staying in the harmonic regime, leading to
\begin{equation}
E = \frac{1}{V_0} \left. \frac{\partial^2 U}{\partial \epsilon^2} \right|_{\epsilon=0} \; ,
\label{Formule_E}
\end{equation}
as an expression of the Young's modulus $E$. $V_0$ defines the volume at equilibrium and $U$ the total energy.
A detailed description of the Young's modulus calculations is given in \cite{submat}.\\

\section{Volume definition based on the electron density}

In order to define nanoobject volume, we start with the average electron density $\rho$ of a bulk material.
This can always be defined as $\rho_{bulk} = Q_{total} / V_0$, where $Q_{total}$ gives the total number of electrons in a cell of volume $V_0$, \textit{e.g.} the conventional unit cell.
For any system the local electron density $n(\vec{r}_i)$ $(i=1..N)$ can be generated in real space at every point $\vec{r}_i$ in a fine uniform 3D mesh of $N$ points in a supercell.  Many DFT codes such as \textit{AIMPRO} already define a real-space 3D mesh to describe the system electron density, and thus for computational efficiency we use the pre-generated mesh in the following analysis. The grid mesh density is sufficiently fine that the final calculated volume is converged to less than 1\% variation (see \cite{submat}).\\

The total number of electrons in the supercell (SC) with known volume $V_{SC}$ is fixed, and can be expressed as the sum of the electron density over all points multiplied by the fractional volume associated with every point,
\begin{equation}
Q_{total} = \frac{V_{SC}}{N} \cdot \sum\limits_{i=1}^N n(\vec{r_i}) \; .
\label{Formule_Q}
\end{equation}
This definition is independent of the type of structure or supercell, for example a bulk calculation or a single-layer nanosheet surrounded by vacuum. 
In order to define nanoobject volume we now introduce an electron density cut-off $c$.  We can find all the points $N_{n>c}$ with electron density $n(\vec{r_i}) > c$. This leads to the number of electrons $Q(c)$ and volume $V(c)$, knowing $V_{SC}$ and the number of grid points $N$,

\begin{eqnarray}
Q(c) &=& \frac{V_{SC}}{N} \cdot  \sum\limits_{i=1}^{N_{n>c}} n(\vec{r_i}) \;,
\label{Formule_Qc} \\
V(c) &=& \frac{N_{n>c}}{N} \cdot V_{SC} \;.
\label{Formule_Vc}
\end{eqnarray}

\begin{figure}
\centering
\includegraphics[width=0.7\linewidth]{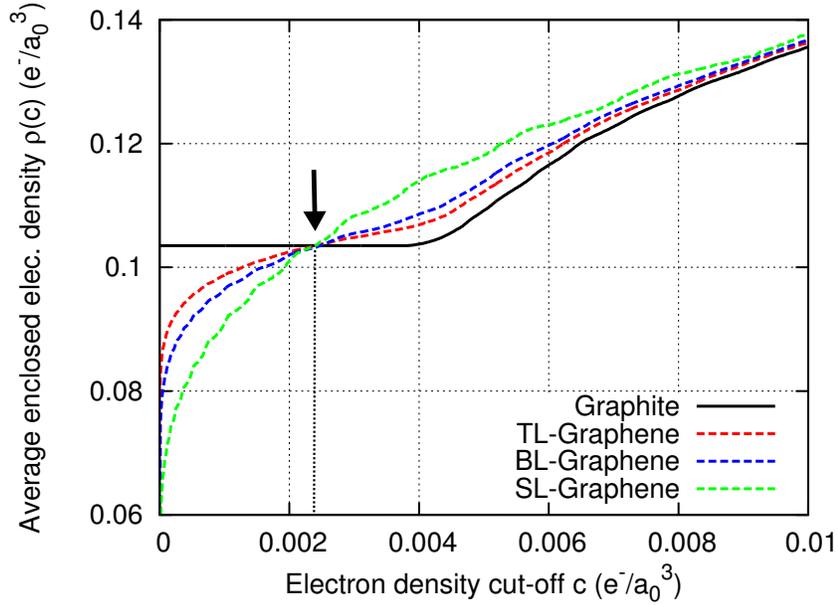}
\caption{Average enclosed electron density $\rho(c)= Q(c)/V(c)$ as a function of the electron density cut-off $c$ for single-layer (SL), bi-layer (BL), tri-layer (TL) graphene and graphite. The arrow indicates the cut-off given by (\ref{Formule_definitioncutoff}), values shown in Table \ref{table_layerYM}.}
 \label{Density}
\end{figure}

We propose to choose this electron density cut-off such that \textit{the resultant nanoobject volume (here nanosheet or nanotube volume) has the same average electron density as the parent (layered) bulk material:}\\ 

\begin{equation}
\rho_{bulk} = \left(\frac{Q(c)}{V(c)}\right)_{nanoobject} = \rho(c)_{nanoobject} \;.  
\label{Formule_definitioncutoff}
\end{equation}

This leads to a new expression for the volume $V(c)=Q(c)/\rho_{bulk}$ where $c$ corresponds to the crossing point of the average electron densities for nanosheet or nanotube and parent layered bulk material, as indicated with an arrow in Figure \ref{Density} for single- (SL), bi- (BL), tri-layer (TL) graphene with graphite. 
In all systems examined here these volumes enclose more than $99.35$ \% of the total electrons in the supercell (see Table \ref{table_layerYM} and \ref{table_CNTs} and \cite{submat}). 
The Young's modulus $E$ can now be expressed using volume $V_{0}(c)$, which is only dependent on the electron distribution and thus takes directly into account the geometry of the structure,
\begin{equation}
E(c) = \frac{1}{V_{0}(c)} \left. \frac{\partial^2 U}{\partial \epsilon^2} \right|_{\epsilon=0} \; .
\label{Formule_Efinal}
\end{equation}

\section{Young's modulus of nanosheets and nanotubes}

Calculated Young's moduli using the new volume definition for single-, bi- and tri-layer graphene are in good agreement with experimental values of $\approx 1$ TPa \cite{Lee2009} (see Table \ref{table_layerYM}).  Since these experimental values assume slabs with graphite inter-layer thickness of $3.35$ \AA, we converted our volumes into equivalent slab thicknesses for comparison. Although we note that the enclosed volumes are in reality not uniform slabs but show surface undulation reflecting the electron distribution in the underlying lattice. The equivalent layer thickness we obtain varies with the number of layers, from $3.31$ \AA\; for SL-graphene converging towards our calculated graphite layer spacing of $3.32(3)$ \AA\; with increasing layer number. This slighly smaller inter-layer distance for graphite results from the chosen pseudo-potentials and the LDA-DFT approach.\\ 
\begin{table}
\centering
\caption{\label{table_layerYM} Calculated in-plane Young's modulus $E$ for different nanosheets and their parent bulk materials. $t$ indicates the single-layer thickness of a slab with equivalent volume to that defined by the electron density cut-off $c$. $N_Q=Q(c)/Q_{total}$ gives the ratio of enclosed electrons compared to the total number of electrons in the supercell.}
\begin{tabular}{l c c c c }
\br
Sheets  & $E(c)$ (TPa) & $c$ (e$^{-}$/a$_0^3$) & $t$ (\AA) & $N_Q$ (\%)\\
\mr
SL-Graphene & 1.059 & 0.00240 & 3.31 & 99.64 \\ 
BL-Graphene & 1.059 & 0.00247 & 3.32 & 99.81 \\
TL-Graphene & 1.058 & 0.00237 & 3.32 & 99.88 \\ 
4L-Graphene & 1.055 & 0.00226 & 3.32 & 99.91 \\
Graphite (bulk) & 1.055 & - & 3.32 & 100.0 \\
\\
SL-BN  &  0.898 & 0.00268 & 3.19 & 99.60 \\ 
BL-BN & 0.891 & 0.00288 & 3.19 & 99.78 \\ 
TL-BN & 0.886 & 0.00277 & 3.19 & 99.86 \\ 
h-BN (bulk) & 0.880 & - & 3.19 & 100.0 \\
\\
SL-WS$_2$ & 0.251 & 0.00290 & 6.14 & 99.89 \\ 
WS$_2$ (bulk) & 0.242 & - & 6.17 & 100.0 \\
\\
SL-MoS$_2$  & 0.222 & 0.00293 & 6.12 & 99.85 \\ 
MoS$_2$ (bulk) & 0.219 & - & 6.14 & 100.0 \\
\\
SL-MoSe$_2$ & 0.188  & 0.00335 & 6.35 & 99.87 \\ 
MoSe$_2$ (bulk) & 0.188 & - & 6.36 & 100.0 \\
\\
SL-MoTe$_2$ & 0.132 & 0.00329 & 6.87 & 99.87 \\ 
MoTe$_2$ (bulk) & 0.132 & - & 6.91 & 100.0 \\
\br
\end{tabular}
\end{table}
We have further calculated the Young's modulus of recently isolated nanosheets \cite{Coleman2011short} using the new volume definition (Table \ref{table_layerYM}). In general single-layer average thicknesses are only slightly smaller than the bulk inter-layer distance, due to the absence of extremely weak inter-layer electron delocalization effects \cite{Charlier1994}.  We obtain good agreement for MoS$_2$ with experiment, the only one of these to be experimentally determined to date to the best of our knowledge ($0.27\pm0.1$ TPa \cite{Bertolazzi2011}, $0.33\pm0.07$ TPa \cite{Castellanos-Gomez2012} compared to our value of $0.222$ TPa).\\

In general the in-plane Young's moduli for nanosheets are similar to their parent bulk material: the in-plane force constants are similar, the out-of-plane interactions are weak, and the single-layer volume is close to that of one bulk layer. This observation makes prediction of nanosheet mechanical properties easier when the Young's modulus of the bulk materials are known.\\ 
Since our calculated averaged graphene layer thickness is close to the interlayer spacing of graphite, this suggests that a $3.35$ \AA \; thick geometric slab is a reasonable approximation to determine pristine graphene volume.  However there are many situations for which the geometric slab model is no longer applicable (for example defective systems such as vacancy-containing graphene), where the new electron density based volume approach proposed here can still be applied.\\

\begin{table}
\centering
\caption{\label{table_CNTs} Axial Young's modulus $E$ calculated for different SWCNTs. $t$ indicates the hypothetical cylinder thickness (brackets indicate completely filled tubes) centred around the SWCNT atom positions, with equivalent volume to that defined by the electron density cut-off $c$. $N_Q=Q(c)/Q_{total}$ gives the ratio of enclosed electrons and $c$ the evaluated electron density cut-off.}
\begin{tabular}{l l c c c c}
\br
 & SWCNT & $E(c)$ (TPa)& $c$ (e$^{-}$/a$_0^3$) & $t$ (\AA) &  $N_Q$ (\%) \\
\mr
(armchair) & (2,2) & 0.642 & 0.00272 & (3.04) & 99.45 \\ 
		   & (3,3) & 1.049 & 0.00255 &  (3.21) & 99.60 \\  
		   & (4,4) &  0.995  & 0.00246 & 3.25 & 99.61 \\ 
		   &  (5,5) & 1.018 & 0.00243 & 3.27 & 99.62 \\  
   		   &  (8,8) & 1.057  & 0.00240 & 3.30 & 99.63 \\ 
   		   &  (10,10) & 1.063 & 0.00238 & 3.31 & 99.64 \\ 
   		   \\
(zigzag)   &  (3,0) & 0.885 & 0.00295 & (3.00) & 99.36 \\ 
		   &  (4,0) &  0.969 & 0.00255 & (3.12) & 99.53 \\ 
		   &  (5,0) & 0.969 & 0.00252 &  (3.20) & 99.61 \\ 
		   &  (6,0) & 1.010 & 0.00247 & 3.23 & 99.61 \\
 		   &  (9,0) & 1.005 & 0.00240 & 3.29 & 99.63 \\ 
           &  (12,0) & 1.028 & 0.00240 & 3.30 & 99.63 \\ 
           &  (17,0) & 1.054 & 0.00236 & 3.31 & 99.64 \\
			\\
(chiral)   &  (4,1) & 1.001 & 0.00244 & (3.17) & 99.60 \\ 
           &  (8,2) & 1.019 & 0.00241 & 3.27 & 99.63 \\
		   &  (8,4) & 1.046 &0.00240 & 3.29 & 99.63 \\ 
           &  (12,6) & 1.054 & 0.00239 & 3.30 & 99.63 \\ 
\mr          
Exp. $E$ & \multicolumn{3}{c}{ \; \; \; \; \; \; \; $\approx$ 1 TPa \cite{Salvetat1999,Yu2000} } &  \\
\br
\end{tabular}
\end{table}

Next the Young's moduli of different SWCNTs have been calculated. In the literature different methods have been applied but all have in common an estimated wall thickness. Using the new volume definition with equal average electron density $\rho(c)$ to graphite, the axial Young's moduli for a range of armchair, zigzag and chiral SWCNTs are summarized in Table \ref{table_CNTs} (for detailed comparison with other theoretical studies see \cite{submat}). The in-plane Young's moduli converge to that of graphite and graphene for larger diameters, and thus lower curvature. The enclosed electron ratio $N_Q$ similarly converges to the graphene value.  However the equivalent wall thickness now varies, and in particular for CNTs with diameters below around $4.7$ \AA \; the CNTs are completely filled (independent of the chirality) \cite{submat}. This agrees with the lower diameter limit for the filling of SWCNTs with water \cite{Cambre2010}.\\

\section{Fermi level dependent Young's modulus and Poisson's ratio of graphene}

\begin{figure}
\centering
\includegraphics[width=0.8\linewidth]{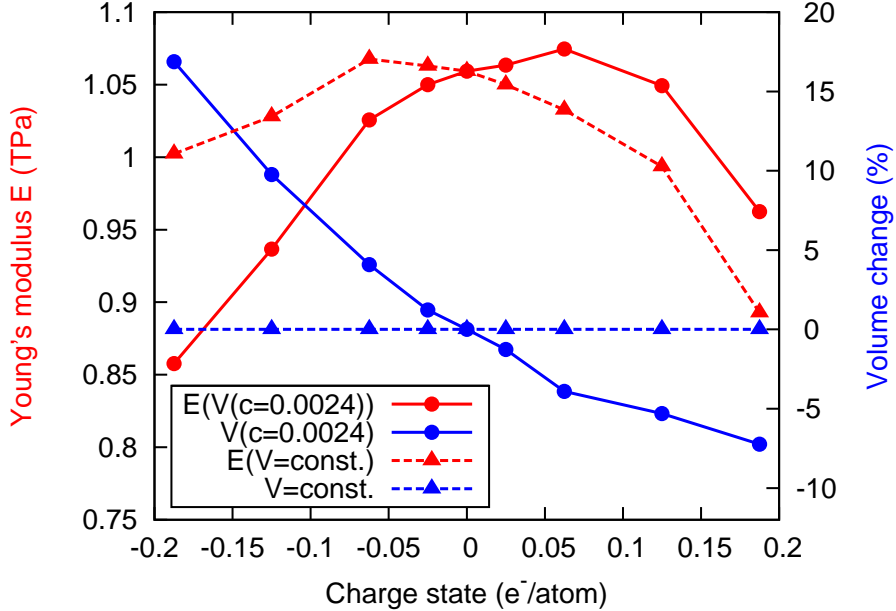}
 \caption{Young's modulus E(V(c=0.0024)) and volume V(c=0.0024) for graphene as a function of doping level/charge state, with volume defined using the graphene electron density cut-off of $c$=0.0024 e$^{-}$/a$_0^3$. The effective Young's modulus E(V=const.) fixing the volume at the charge neutral value shows the modulus variation with charge state due purely to changes in bond strength.}
 \label{Graphene_fermi}
\end{figure}

Defining volume via a well-defined cut-off in the system electron density has further conceptual implications. For example, varying the Fermi level can change the Young's modulus by depopulating bonding states or populating anti-bonding states, softening the bond spring constants of the system and hence the $\partial^{2} U / \partial \epsilon^{2}$ term of (\ref{Formule_Efinal}). However since volume is now defined in terms of a cut-off defined for the system electron density in equilibrium, the enclosed volume will now also be Fermi level dependent. This means that changes in $1 / V_0(c)$ in (\ref{Formule_Efinal}) can also modify the Young's modulus. The current approach includes both of these effects for the first time in the literature. Figure \ref{Graphene_fermi} shows the calculated effect of varying the Fermi level on the Young's modulus of graphene. Over moderate doping levels ($\pm0.0625$ e$^-$/atom), a classical fixed volume model would suggest a gradual drop in Young's modulus as the system becomes more positive (E(V=const.)). However this modulus trend is actually inverted once the corresponding volume decrease is included. Such complex doping-dependence of mechanical properties is not accessible with classical geometrical slab or sphere volume models.\\ 

This new volume definition also enables access to other mechanical properties such as the out-of-plane Poisson's ratio for surface dominated nanoobjects, since it is possible to calculate the volume and hence an equivalent thickness change as the sample is strained. The Poisson's ratio is constant for small strains, and we have taken the average for six strained/compressed cases (see \cite{submat}). For graphene we find the in-plane Poisson's ratio to be $\nu_{12}=0.20$ and for the first time we also calculate the out-of-plane value to be $\nu_{13}=0.015$, using the graphene electron density cut-off $c$=0.0024 e$^{-}$/a$_0^3$ \cite{submat}. Our calculated Poisson's ratios for graphite ($\nu_{12}=0.21$, $\nu_{13}=0.00$) and $\nu_{12}$ for graphene are in good agreement with literature values \cite{Scarpa2009}.\\
We note that for carbon based ``all surface" systems such as single-layer graphene or SWCNTs an electron density cut-off around $0.0024$ e$^{-}$/a$_0^3$ delivers an accurate mechanical volume description with a very stable and very high ratio of enclosed electrons of more than $99.5$ \% (see also \cite{submat}). To apply such a universal material cut-off value to a broader range of structures such as nanoribbons and organic molecules would significantly extend the utility of this volume definition, and will be the subject of a future publication.\\

\section{Conclusion}
To summarise, we propose a new definition of mechanically active volume applicable to nanoobjects dervied from layered bulk materials, using a volume chosen such that the average electron density of the nanoobject matches that of the parent bulk material. This definition is geometry independent, transferable, invokes no empirical parameters and can be implemented in all standard DFT approaches. It correctly extrapolates between individual nanoobjects and bulk systems. Since both experimental and theoretical derivation of Young's modulus require a volume definition, the same calculated volumes can be applied to both. Based on this one general volume definition, for the first time consistent and comparable values for Young's moduli of various new nanosheets and single-wall carbon nanotubes have been calculated. All values show good agreement with the parent bulk in-plane Young's modulus. This can be really stated for the first time, as the calculations are based on a transferable underlying method.
In addition this new approach allows study of systems whose volume varies, for example by shifting the Fermi level. It can be easily applied to nanostructures containing defects such as vacancies, which will locally modify the electron density distribution and hence volume. This volume definition could also be applied in other systems where nanoscale volume is needed such as the definition of internal porosity for metal-oxide frameworks.\\

\ack
PW, VVI and CPE acknowledge project NANOSIM-GRAPHENE n$^{\circ}$ANR-09-NANO-016-01 funded by the French National Agency (ANR). We thank the CCIPL and COST Project MP0901 NanoTP for support. MJR thanks the Swedish Foundation for Strategic Research for financial support. PRB thanks the CNRS for financial support.\\

\vspace*{2cm}

\textit{For references, please see at the end of the document.}

\newpage

\title[Mechanical properties of nanosheets and nanotubes ...]{Supplementary materials\\ \small{\textit{J. Phys.: Condens. Matter, 25 (15) 155302, 2013 }}}

\section*{Literature approaches used to calculate Young's modulus}

All theoretical calculations of Young's modulus of graphene to date have either determined the force or the total system energy change, under applied strain.  These have been determined using a range of different levels of theory, including density functional theory under LDA \cite{Liu2007,Wagner2011a}, GGA \cite{Faccio2009} and B3LYP \cite{Zeinalipour-Yazdi2009}, Hartree-Fock formalism \cite{VanLier2000}, tight-binding formalism \cite{Hernandez1998}, empirical force constant models such as REBO \cite{Lu1997,Zhang2011a} or Brenner potentials \cite{Gupta2005}, and empirical lattice dynamics models \cite{JianPing1997}.  Despite the variety of techniques, the calculated values when choosing the same sample slab or cylinder thickness generally fall in a similar range.

\section*{Method to calculate the Young's modulus}

The Young's Modulus $E$ is defined as stress $\sigma$ over strain $\epsilon$,
\begin{equation}
E = \sigma / \epsilon \;.
\end{equation} 
The strain can be defined as the fractional change in length $\epsilon=\Delta l / l_0$ along the direction of the applied strain, where $l_0$ defines the length at equilibrium. The stress is defined as the force $F$ per surface area $A$, $\sigma = F/A$. When applying a strain the induced change in energy $\Delta U(\epsilon)$ can be expressed in terms of the total energy $U(\epsilon)$ and the equilibrium total energy $U(0)$ of the system. Using the first derivative of the energy we can write the force as

\begin{equation}
F = - \frac{\partial U(\epsilon)}{\partial l} = - \frac{\partial U(\epsilon)}{\partial l} \frac{\partial l}{\partial \epsilon \; l_0} = -\frac{\partial U(\epsilon)}{\partial\epsilon \; l_0 } \;.
\end{equation}
 $U(\epsilon)$ can be expanded in the form of a $n$th order polynomial,
\begin{equation}
U(\epsilon) = a \epsilon +b \epsilon^2 +c \epsilon^3 +... + C \; .
\end{equation}

\begin{figure}
\centering
\includegraphics[width=9cm]{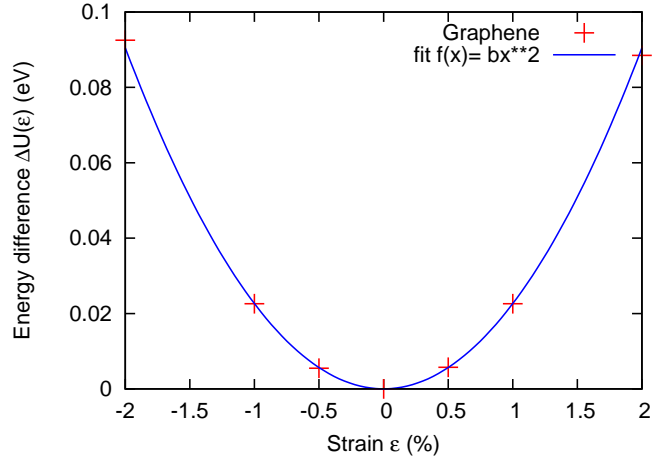}
 \caption{Total energy change for graphene (supercell with 8 carbon atoms) as a function of applied strain, with an associated quadratic fit.}
 \label{fit_graphene}
\end{figure}

For small strains (typically $< 5$\%) we are in the harmonic regime and $U(\epsilon)$ can be simplified to a quadratic function ($a,c,.. \approx 0 $).  An example is given in Figure \ref{fit_graphene}, where the energy difference $\Delta U(\epsilon)$ (\ref{deltaU}) of in-plane strained graphene with a quadratic fit is shown.
\begin{eqnarray}
\Delta U(\epsilon) &=& U(\epsilon)- U(0) \nonumber \\
&=&a \epsilon +b \epsilon^2 +c \epsilon^3 +...\approx b \epsilon^2 \;.
\label{deltaU}
\end{eqnarray}

The Young's modulus $E$ can now be written in a more accessible form using (1), (2), (3) and (4), 
\begin{equation}
E = \frac{F}{A_0 \epsilon} = \frac{\partial U(\epsilon)}{\partial \epsilon} \frac{1}{ \;l_0 \; A_0 \; \epsilon} = \frac{2 b \epsilon}{l_0 A_0 \epsilon }  = \frac{2 b}{V_0}
\end{equation}
where $V_0$ defines the equilibrium volume.
Alternatively, instead of a polynomial expansion, $U(\epsilon)$ can be developed as a Taylor Series \cite{Zeinalipour-Yazdi2009} 
\begin{equation} 
 U(\epsilon) = \left. \frac{\partial U}{\partial \epsilon} \right|_{\epsilon=0} \epsilon + \frac{1}{2} \left. \frac{\partial^2 U}{\partial \epsilon^2} \right|_{\epsilon=0} \epsilon^2 + \frac{1}{6} \left. \frac{\partial^3 U}{\partial \epsilon^3} \right|_{\epsilon=0} \epsilon^3 + ... \,.
\end{equation}

The two approaches are equivalent and the Young's modulus in the regime of small deformations can finally be written in the form of (\ref{Formule_E2}), including the second derivative of the energy and the volume $V_0$.
\begin{equation}
E = \frac{1}{V_0} \left. \frac{\partial^2 U}{\partial \epsilon^2} \right|_{\epsilon=0} = \frac{2 b}{V_0} \; .
\label{Formule_E2}
\end{equation}
This approach has been used successfully by other groups \cite{Zeinalipour-Yazdi2009,Gupta2005,Hernandez1998} and is the basis for the Young's modulus calculations presented here.\\ 

\begin{table}[h!]
\centering

\caption{Calculated (in-plane) Young's modulus $E_{{(}100{)}}$ and average electron density $\rho_{bulk}$ for different layered bulk materials. For comparison some literature values have been recalculated (marked with *) from the given elastic coefficients using $E_{(100)} = \frac{1}{s_{11}} = \frac{(c_{11}+2c_{12})(c_{11}-c_{12})}{c_{11}+c_{12}}$.}
\label{table_bulkYM}

\begin{tabular}{l c c c c}
\br
Bulk & \multicolumn{2}{c}{This work} & Theory & Experiment \\
Material & E (TPa) & $\rho_{bulk}$ (e$^-$/a$_0^3$) & E (TPa) & E (TPa)\\
\mr
Graphite & 1.055 & 0.104 & 1.041$^*$\cite{Qi2010} & 1.060 \cite{Kis2003} \\
& & & 1.029$^*$\cite{Kudin2001} & 1.02$\pm 0.03$\cite{Blakslee1970} \\

h-BN &  0.880      &   0.104   &   0.900$^*$\cite{Ohba2001}  &  0.753$^*$\cite{Bosak2006}\\ 
	&			  &		   &  0.810 \cite{Kudin2001}	& 0.700$^*$\cite{Duclaux1992} \\	

WS$_2$  & 0.242 & 0.296 & 0.238$^*$\cite{Khare2008} &  0.150 \cite{Sourisseau1991}\\
MoS$_2$  & 0.219 & 0.206 & 0.202$^*$\cite{Khare2008} & 0.238 \cite{Feldman1976}\\
MoSe$_2$ & 0.188 & 0.278 & 0.178$^*$\cite{Khare2008}  & - \\
MoTe$_2$ & 0.132 & 0.295 & 0.124$^*$\cite{Khare2008} & - \\

\br
\end{tabular}
\end{table}

Pure bulk (in-plane) Young's moduli have been calculated for different layered materials, shown in Table \ref{table_bulkYM}. To obtain the energy curve the structures were relaxed at $\epsilon$ = $\pm$0.5, $\pm$1, $\pm$2 \% fixed strain along the ${[}100{]}$ direction (in the basal planes). In these cases the bulk volume $V_0$ is clearly defined by the dimension of the supercell (SC) used (i.e. $V_{0}=V_{SC}$).  These calculations do not rely on any nanoobject volume definition but use instead well-defined bulk volumes, and can therefore be used to validate the DFT \textit{AIMPRO} approach utilised.  The agreement between our results and literature show that our potential energy curves are accurately calculated and the error from using a harmonic fit is negligible (in general less than $\pm3$ \%).\\ 

These results are in general in very good agreement with the literature, although some small differences between theoretical and experimental values are to be expected.  This is firstly due to the LDA-DFT approach, where bonding tends in general to be overestimated, which can lead to slightly higher in-plane forces and thus lead to sightly higher values for the Young's moduli.  Secondly the experimental values can be influenced by factors such as sample purity, internal defects and grain size (for example in \cite{Duclaux1992} the BN sample is pyrolitic). Additionally the measurements are often indirect (e.g. the BN values are taken from thermal conductivity studies \cite{Duclaux1992} and x-ray scattering measurements \cite{Bosak2006}). Thus a difference of up to $\sim$ 20 \% between the theoretical and experimental values seems not unreasonable.\\

\section*{Defining the volume based on the average electron density}

The variation of the ratio of enclosed electrons $N_{Q}(c)$ (\ref{Formule_Qc}) and Volume V(c) (\ref{Formule_Vc}) with electron density cut-off $c$ is shown in Figure \ref{Density}(a) and (b), for single- (SL), bi- (BL), tri-layer (TL) graphene and graphite.
\begin{eqnarray}
N_{Q}(c)&=& Q(c)/Q_{total} = \frac{V_{SC}}{N} \cdot  \sum\limits_{i=1}^{N_{n>c}} n(\vec{r_i}) / Q_{total} \;,
\label{Formule_Qc} \\
V(c) &=& \frac{N_{n>c}}{N} \cdot V_{SC} \;.
\label{Formule_Vc}
\end{eqnarray}
For graphite a cut-off of $c$ = 0.0038 e$^-$/a$_0^3$ already includes all points in the cell (see Figure \ref{Density}(c)), and hence for cut-offs less than this there is no change in included volume or electrons.  In contrast, for graphene the volume $V(c)$ increases rapidly as the cut-off drops, which can be understood as the surface charge smears out from the graphene into the vacuum around it.  Even grid points with $n(\vec{r})$ close to zero can be found.  However this smeared out charge represents only a very small fraction of the total number of electrons (see Figure \ref{Density} (a)).  For example at the minimal graphite cut-off of $c$ = 0.0038 e$^-$/a$_0^3$, already more than 99 \% of the total number of electrons per atom in graphene are included. In general the enclosed volume is relatively insensitive to the precise cut-off value, since the electron density in the region surrounding the volume cut-off drops away rapidly. For example, increasing $c$ from our obtained graphene cut-off value of $c$=0.0024 e$^-$/a$_0^3$ to 0.003 e$^-$/a$_0^3$ only decreases the enclosed volume by 4.5\%, with $N_Q$ = 99.52 \% still.\\
Figure \ref{Density} (d) shows that the calculated nanoobject volume converges rapidly with the grid mesh used to define the electron density. The grid mesh is given by the plane wave energy cut-off in our calculations.  The volume is already converged for a plane-wave cut-off of 100 Ha. In our graphene and CNT calculations a plane wave energy cut-off of 150 Ha is used. As example for SL-graphene we used a
48$\times$54$\times$180 grid (150 Ha) in a 7.99$\times$9.23$\times$30.00 a$_0^3$ orthorhombic supercell (see Table \ref{table_layerYM2} and Figure \ref{Density} (d)). For SL-graphene calculations an energetically well converged 24$\times$24$\times$1 k-point grid has been used. 

\begin{figure}
\centering
\subfigure [] {\includegraphics[height=7.5cm,angle=-90]{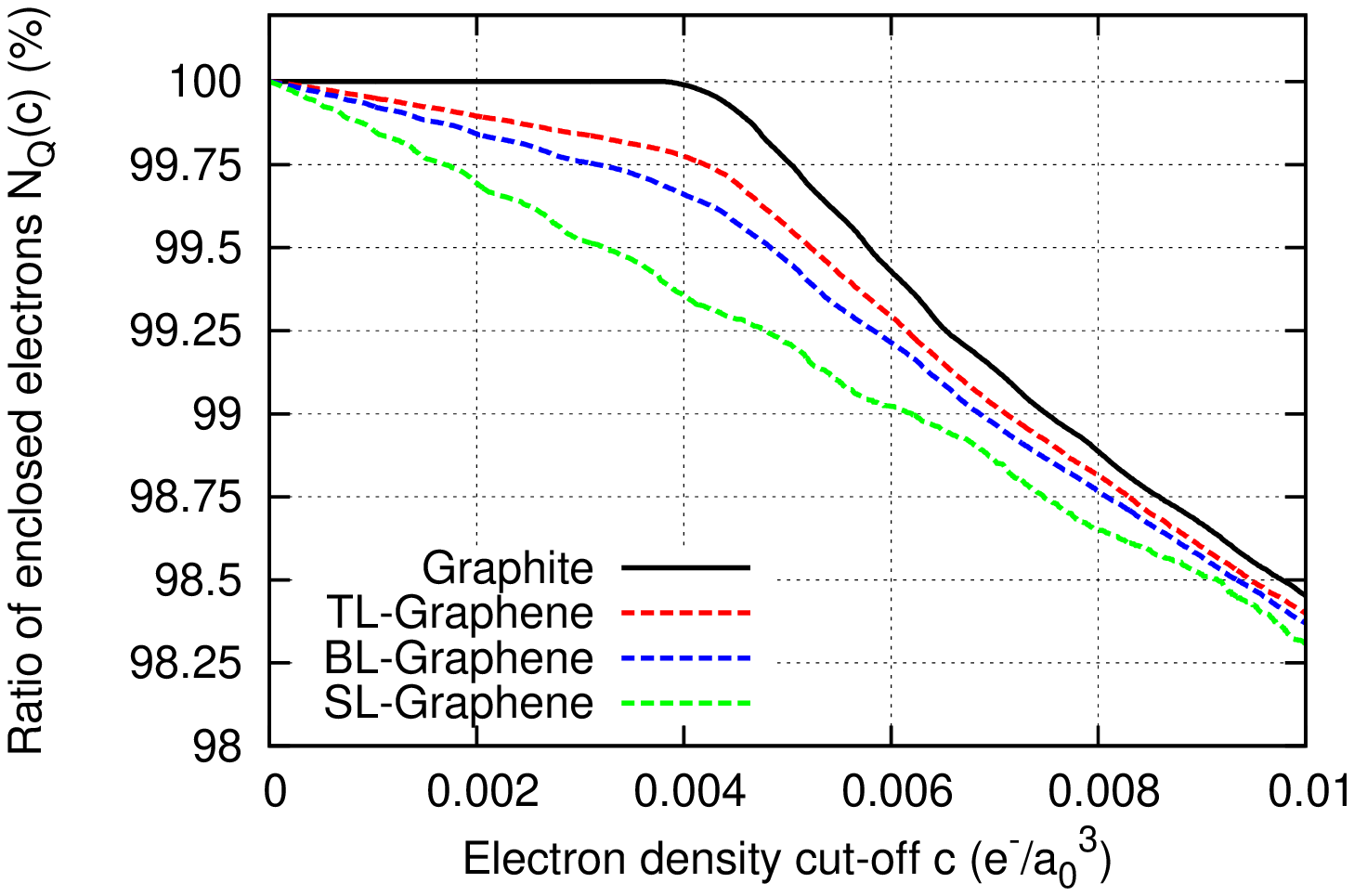}}\hspace*{0.3cm}
\subfigure [] {\includegraphics[height=7.5cm,angle=-90]{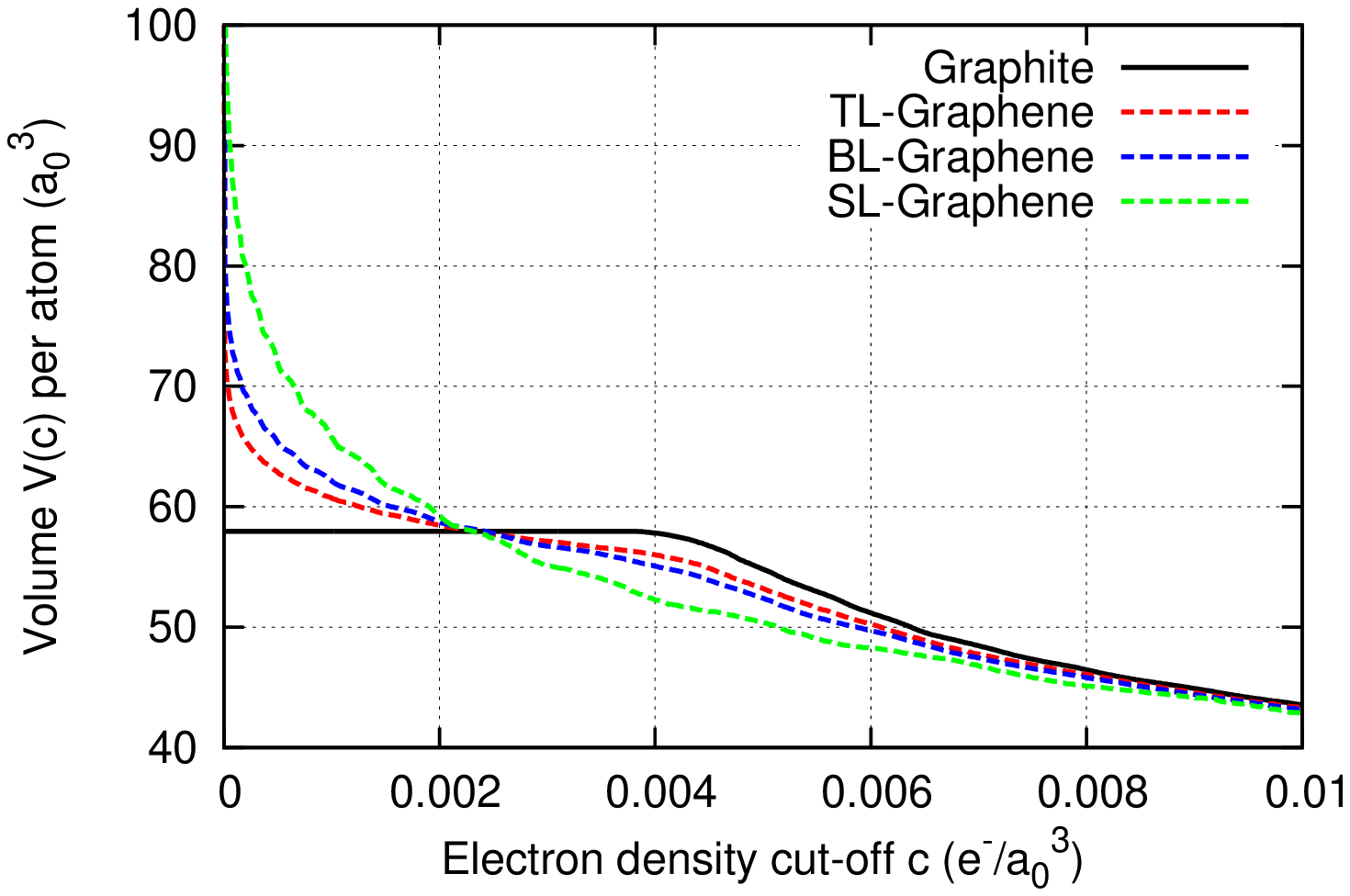}}\\
\subfigure [] {\includegraphics[width=7.2cm]{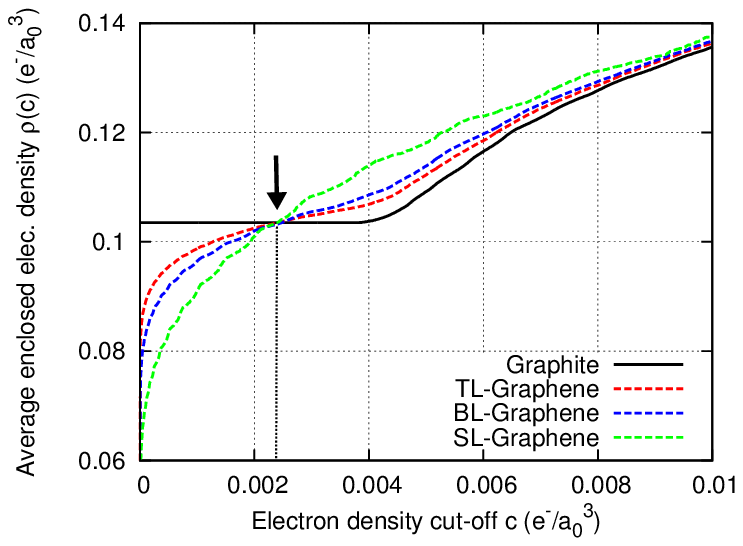}} \hspace*{0.3cm}
\subfigure [] {\includegraphics[width=7.5cm]{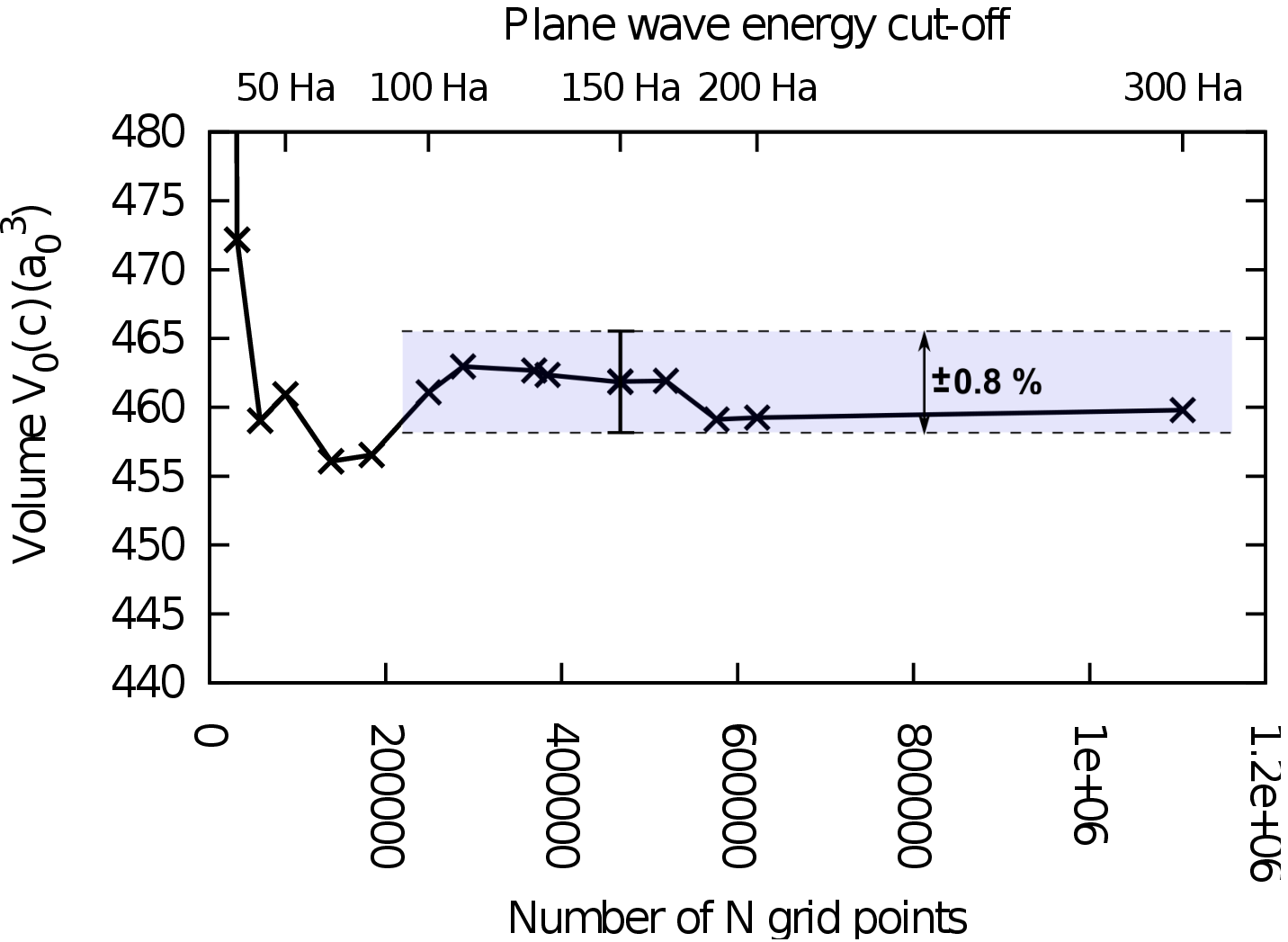}}\\
\subfigure [] {\includegraphics[width=7.5cm]{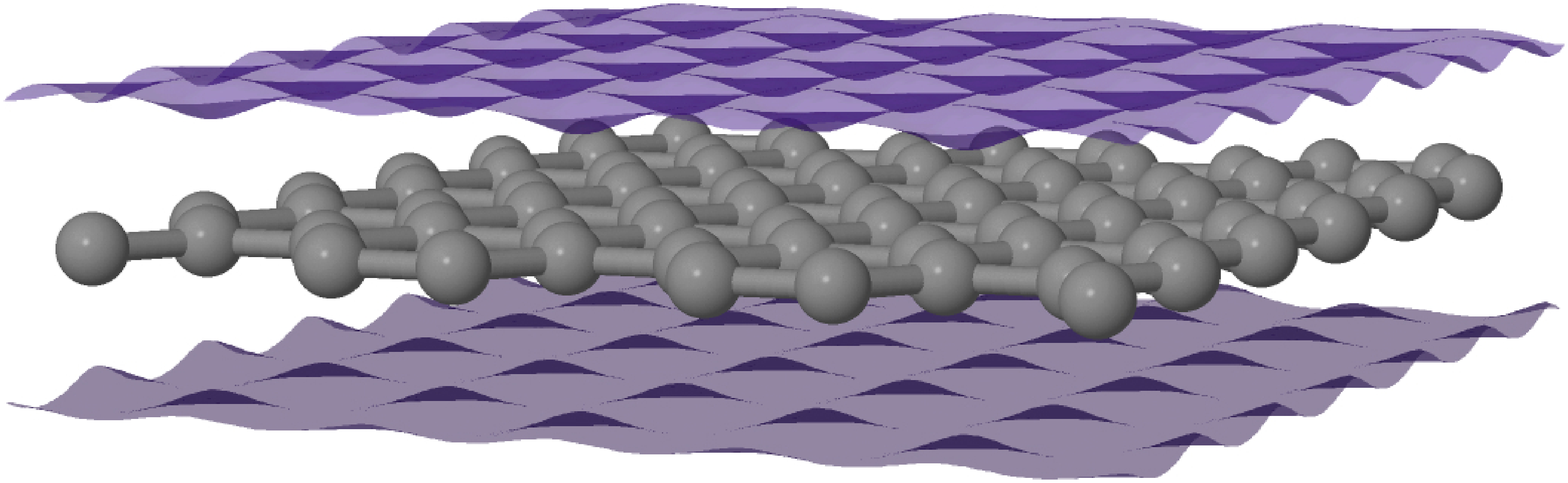}} \hspace*{2.0cm}
\subfigure [] {\includegraphics[width=4.5cm]{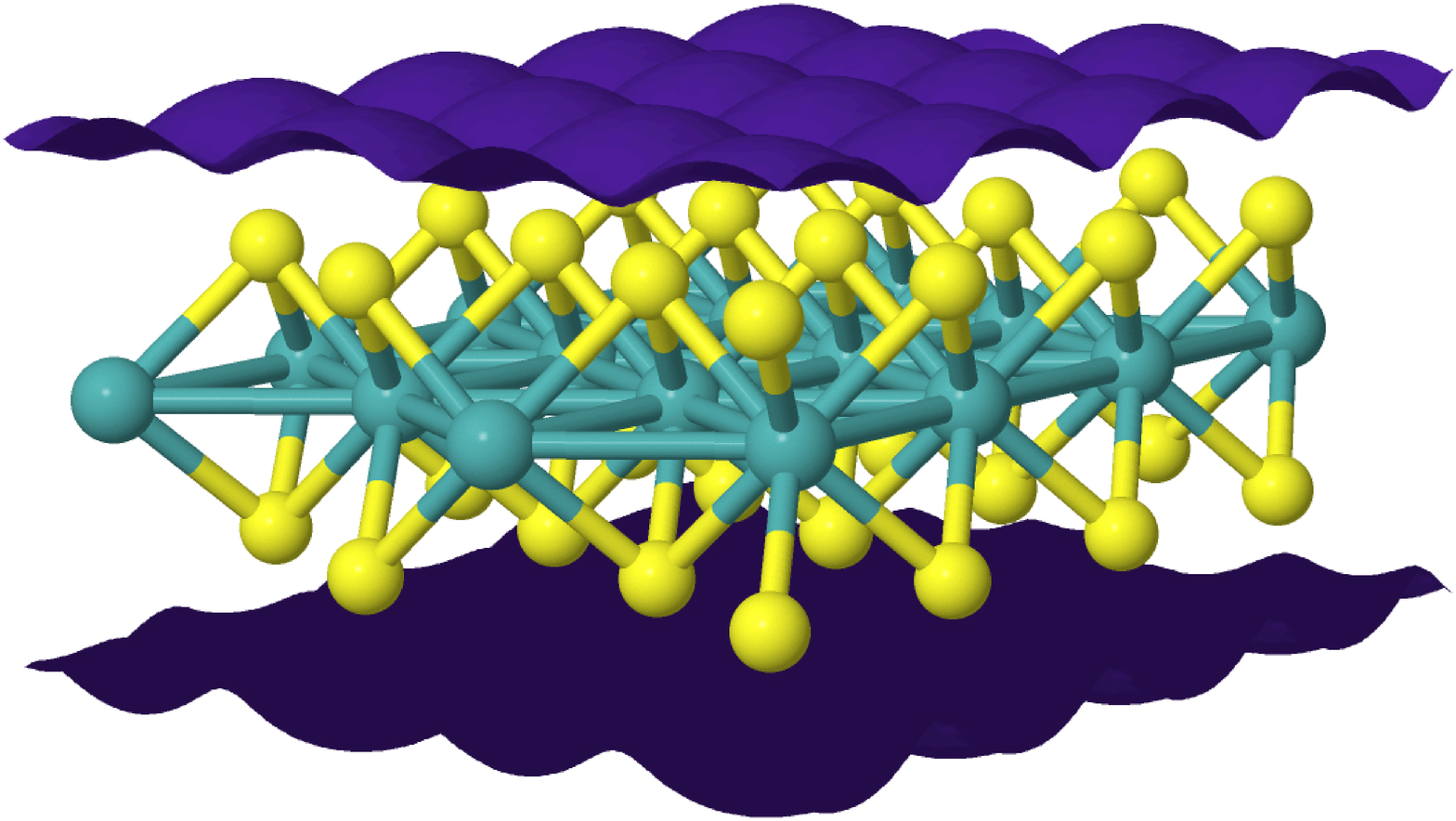}}

\caption{Variation of ratio of enclosed electrons $N_{Q}(c)$ (a), and volume per atom $V(c)$ (b) as a function of electron density cut-off for single-layer (SL), bi-layer (BL), tri-layer (TL) graphene and graphite. (c) Shows the resultant enclosed electron density $\rho(c)= Q(c)/V(c)$. The arrow indicates the cut-off $c$. In (d) the variation of confined volume for SL-graphene (c = 0.0024 e$^{-}$/a$_0^3$) as a function of plane wave energy cut-off is plotted, showing that the volume is essentially converged around 100 Ha (maximum error of less than $\pm$ 0.8 \% taking 150 Ha, \textit{i.e.} 466560 grid points).  Schematic perspective images of volumes, visualized using the iso-surface of the evaluated cut-off $c$, are shown in (e) for SL-graphene and (f) for a MoS$_2$ single-layer nanosheet. Atom colours, grey: C, yellow: S, cyan: Mo.}
 \label{Density}
\end{figure}

\begin{table}[h!]
\centering

\caption{Parameters from the nanosheet calculations (Table 1 in the article). $N_{atoms}$ gives the number of atoms in the supercell, $V_0(c)$ the volume calculated with the given cut-off $c$, and $a,b$ the relaxed in-plane orthorhombic lattice parameters in equilibrium. (In direction $c$ there was always enough space to avoid interactions. An orthorhombic supercell has always been used.)}
\label{table_layerYM2}

\begin{tabular}{l c c c c c }
\br
Sheets  & $N_{atoms}$ & $c$ (e$^{-}$/a$_0^3$) & $V_0(c)$ (a$_0^3$) & $a$ (a$_0$) & $b$ (a$_0$)\\
  
\mr
SL-Graphene & 8 & 0.00240 & 461.82 & 7.996 & 9.232\\
BL-Graphene & 16 & 0.00247 & 926.024 & 7.997 & 9.234\\
TL-Graphene & 24 & 0.00237 & 1390.69 & 7.997 & 9.234\\
4L-Graphene & 32 & 0.00226 & 1855.15 & 7.997 & 9.234 \\  
\\
SL-BN  &  8 & 0.00268 & 461.57 & 8.139 & 9.397 \\
BL-BN & 16 & 0.00288 & 921.68 & 8.143 & 9.402 \\
TL-BN & 24 & 0.00277 & 1383.85 & 8.143 & 9.407\\ 
\\
SL-WS$_2$  & 12 & 0.00290 & 1426.08 & 10.316 & 11.911\\
SL-MoS$_2$  & 12 & 0.00293 & 1426.08 & 10.338 & 11.938 \\
SL-MoSe$_2$ & 12  & 0.00335 & 1571.81 & 10.654 & 12.299 \\
SL-MoTe$_2$ & 12 & 0.00329 & 1962.38 & 11.451 & 13.207\\
\br
\end{tabular}
\end{table}

\pagebreak

\section*{Case study SWCNTs}
\begin{figure}[h!]
\centering
\subfigure [] {\includegraphics[width=8cm]{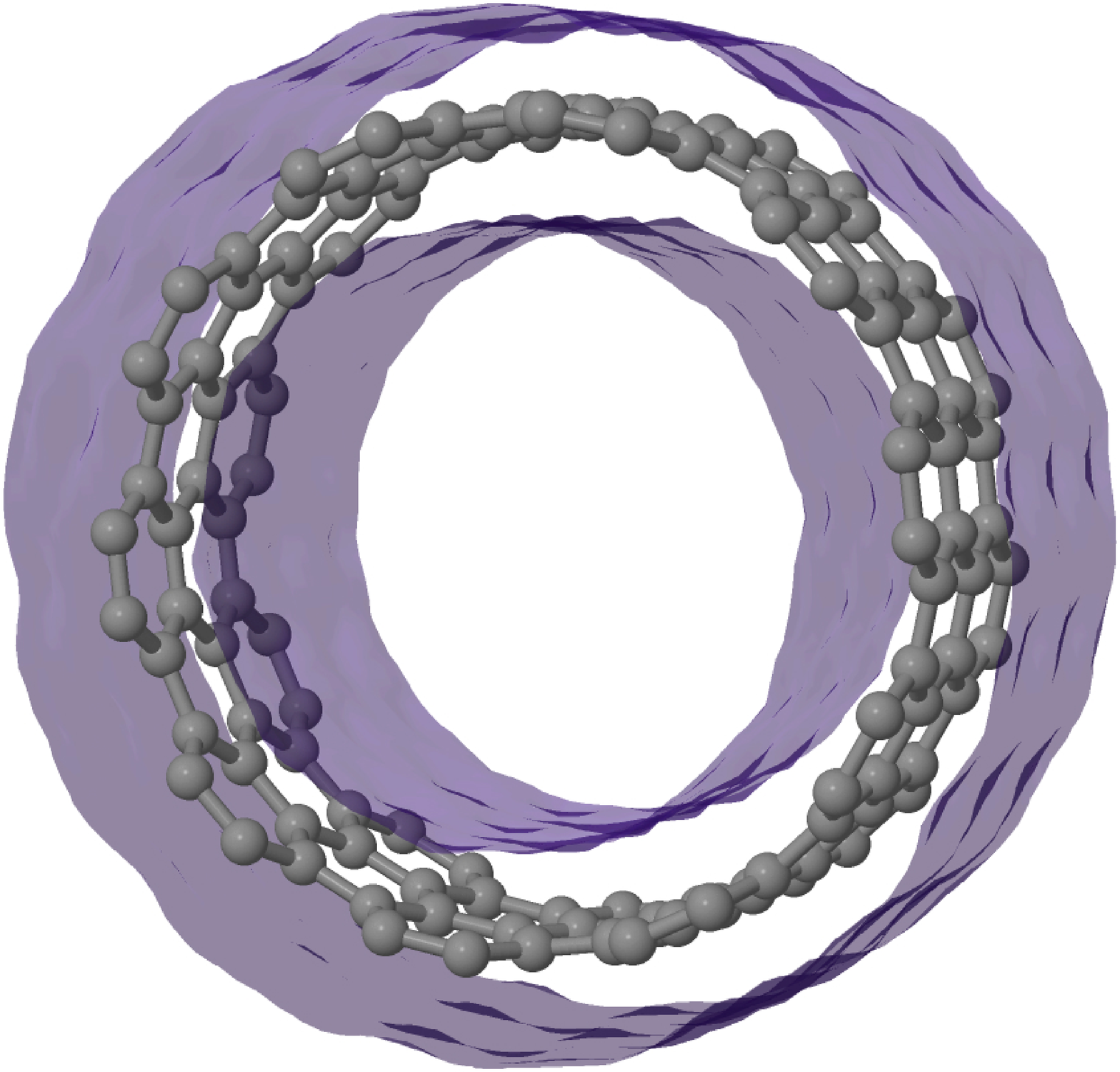}} \hspace*{2cm}
\subfigure [] {\includegraphics[width=4.5cm]{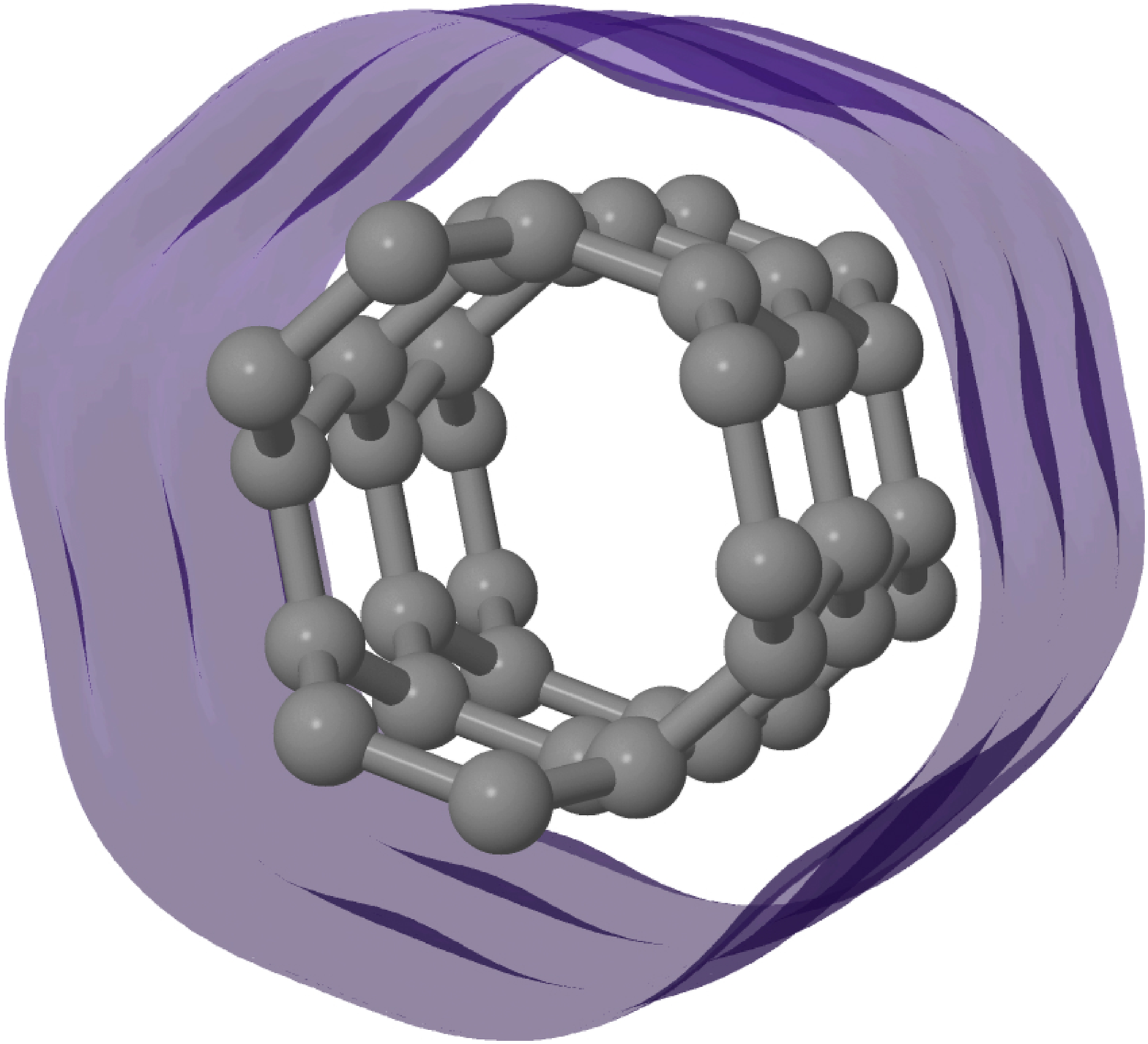}}
\caption{Schematic perspective images of CNT volumes visualized using the iso-surface of the evaluated cut-off $c$, (a) segment of (10,10) SWCNT and (b) segment of (3,3) SWCNT (note the lack of interior cavity in (b)).}
 \label{Volumes}
\end{figure}

\begin{figure}[h!]
\centering
\subfigure [] {\includegraphics[width=7.5cm]{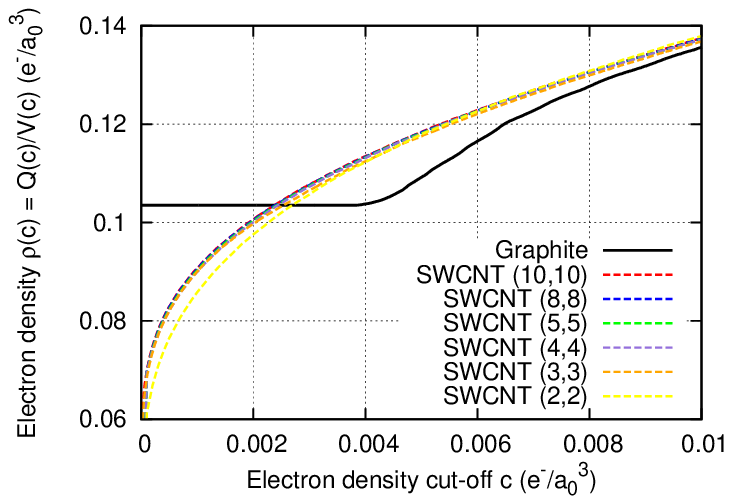}}\hspace*{0.3cm}
\subfigure [] {\includegraphics[width=7.5cm]{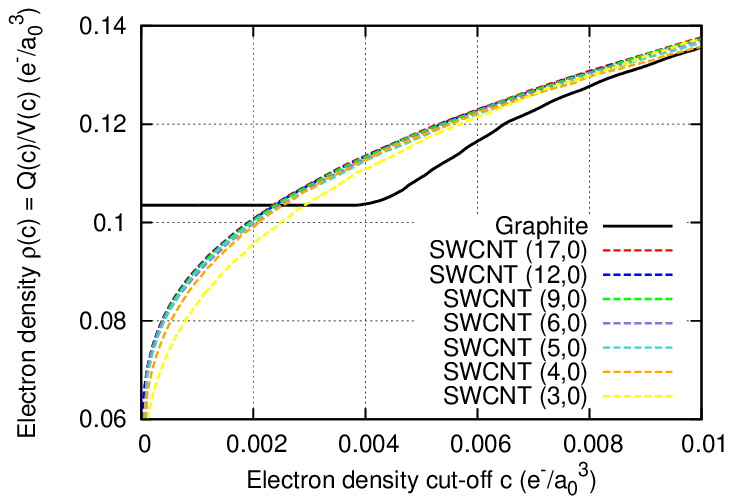}} \\
\subfigure [] {\includegraphics[width=7.5cm]{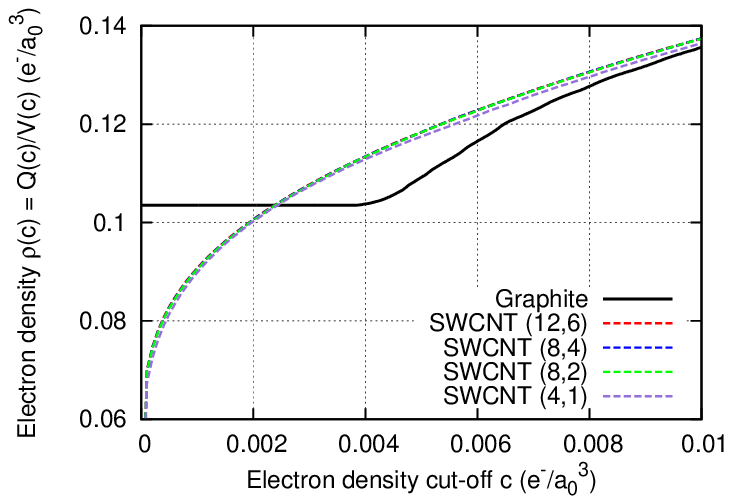}}\hspace*{0.3cm}
\subfigure [] {\includegraphics[width=7.5cm]{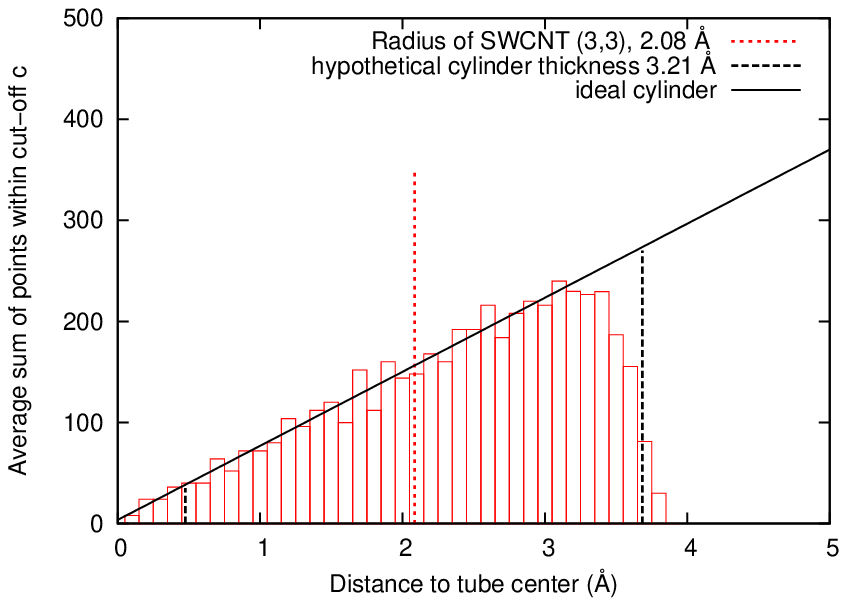}}
\caption{Comparing the average electron density $\rho(c)$ of (a) armchair SWCNTs, (b) zigzag SWCNTs and (c) chiral SWCNTs with graphite as a function of the electron density cut-off $c$. (d) Radial distribution of real space points within the electron density cut-off as a function of distance from the nanotube axis (averaged along tube axis, grid of 30 steps along the tube).  Dotted red line marks distance of atom centres from the axis. The solid black line shows the number of points at a given radius expected for an ideal cylinder, and dotted black lines show limits of a hypothetical cylinder with inner and outer radii evenly spaced around the atom positions, with same total volume as that obtained using the electron density cut-off.}
 \label{CNTDensity}
\end{figure}

Initial experimental studies found axial Young's Modulus values for SWCNTs of around $1.25$ TPa, although more recent studies have found values closer to $1$ TPa.  Our calculated values, along with other theoretical studies are in good agreement with this, shown in Table \ref{table_CNTs}.  This table also shows that Young's modulus becomes diameter independent for tube diameters larger than $\sim$ 4 \AA \; .  
Isosurfaces associated with our calculated nanoobject volume for the (3,3) and (10,10) carbon nanotubes are shown in Figure \ref{Volumes}. In Figure \ref{CNTDensity} the average electron density $\rho(c)$ of the armchair, zigzag and chiral SWCNTs as function of the cut-off $c$ are plotted. Additionally in Figure \ref{CNTDensity} (d) an example of a radial distribution of the grid points is given.\\ 

\vspace*{2cm}

\pagebreak

\begin{table}[h!]
\centering
\caption{Axial Young's modulus calculated for different SWCNTs. For the potential energy curve the SWCNTs are strained  $\pm$0.5, $\pm$1, $\pm$2 \% along their axis. Additionally the electron density cut-off $c$ and the tube diameter $d$ are indicated.\\ (ac: armchair, zz: zigzag, $d$ is calculated out of 3 carbon atomic positions lying in the same plane perpendicular to the tube axis)}
\label{table_CNTs}

\begin{tabular}{l l c c c | c }
\br
 &  & \multicolumn{3}{c}{This work} & Theory \\
 & SWCNT & $E(c)$ (TPa)& c (e$^-$/a$_0^3$) & $d$ (\AA)  & $E$ (TPa) \\
\mr
(ac) & (2,2) & 0.642 & 0.00272 & 2.79 &- \\
		   & (3,3) & 1.049 & 0.00255 & 4.17 &- \\ 
		   & (4,4) &  0.995  & 0.00246 & 5.48 & 0.96 \cite{Akdim2003}  \\
		   &  (5,5) & 1.018 & 0.00243 & 6.82 & 1.11 \cite{Gupta2005} \\ 
		   &        &        &   &  & 0.971 \cite{Lu1997} \\   	   
   		   &  (8,8) & 1.057  & 0.00240 &10.83 & 0.99 \cite{Akdim2003}\\
   		   &        &       &  & & 0.979/1.008 \cite{Ogata2003}\\    	   
   		   &  (10,10) & 1.063 & 0.00238 & 13.46 & 1.23 \cite{Gupta2005}\\
   		   &          &         & & & 0.99 \cite{Akdim2003}\\ 
		   &		  &    		& & & 1.24 \cite{Hernandez1998} \\  
   		   &          &         & & & 0.972 \cite{Lu1997} \\
   		   \mr
(zz)   &  (3,0) & 0.885 & 0.00295 & 2.60 &  - \\
		   &  (4,0) &  0.969 & 0.00255 & 3.34 & 0.84 \cite{Akdim2003} \\
		   &  (5,0) & 0.969 & 0.00252 & 4.04 & - \\
		   &  (6,0) & 1.010 & 0.00247 & 4.79 & 0.96 \cite{Akdim2003} \\
 		   &  (9,0) & 1.005 & 0.00240 & 7.05 & 1.16 \cite{Gupta2005}\\
		   &        &          & & & 0.974/1.017 \cite{Ogata2003} \\
           &  (12,0) & 1.028 & 0.00240 & 9.37 & - \\
           &  (17,0) & 1.054 & 0.00236 & 13.20 &  1.227 \cite{Gupta2005}\\
			\mr
(chiral)   &  (4,1) & 1.001 & 0.00244 & 3.73 & -\\
           &  (8,2) & 1.019 & 0.00241 & 7.21 & 0.974 \cite{Lu1997} \\
		   &  (8,4) & 1.046 & 0.00240 & 8.29 & 1.176 \cite{Gupta2005}\\
           &  (12,6) & 1.054 & 0.00239 & 12.39 & 1.20 \cite{Gupta2005}\\
\mr        
Exp. $E$ & \multicolumn{4}{c}{\;\;\;\; $\approx$ 1\;TPa \cite{Wong1997,Salvetat1999,Yu2000,Coleman2006}, 1.25\;TPa \cite{Krishnan1998} }   \\
\br
\end{tabular}
\end{table}

\vspace*{8cm}
\newpage
\section*{Fermi level dependent Young's modulus}

Table \ref{Fermi_level} gives a detailed overview of the results obtained when changing the Fermi level / charge state of a single graphene layer. We note that the enclosed electron ratio $N_Q$ varies very little, demonstrating the reliability of the constant-cut-off $c=0.0024$ e$^-$/a$_0^3$.\\

\begin{table}[h!]
\centering

\caption{Calculations of single-layer graphene. For all volumes $V(c)$ a cut-off $c=0.0024$ e$^-$/a$_0^3$ has been used. $N_Q=Q(c)/Q_{total}$ gives the ratio of enclosed electrons. $a,b$ are the relaxed in-plane orthorhombic lattice parameter using a supercell with 8 carbon atoms.}
\label{Fermi_level}

\begin{tabular}{l c c c c }
\br
charge state (e$^{-}$/atom) & $V(c)$ (a$_0^3$) & $a$ (a$_0$) & $b$ (a$_0$) & $N_Q$ (\%) \\
\mr
-0.1875 & 539.75 &  8.258 & 9.535 & 99.61 \\
-0.125 & 506.87 & 8.143 & 9.403 & 99.61 \\
-0.0625 & 480.66 & 8.056 & 9.302 & 99.62 \\
-0.025 & 467.49 & 8.016 & 9.255 & 99.63 \\
0.0 & 461.82 & 7.996 & 9.232 & 99.64 \\
+0.025 & 455.95 & 7.985 & 9.219 & 99.65 \\
+0.0625 & 443.78 & 7.979 & 9.215 & 99.63 \\
+0.125 & 437.30 & 8.001 & 9.237 & 99.66 \\
+0.1875 & 428.41 & 8.054  & 9.297 & 99.64 \\
\br
\end{tabular}
\end{table}

\newpage
\section*{Poisson's ratio of graphene}

When calculating the out-of-plane Poisson's ratio of graphene, the ratio of enclosed electrons $N_Q$ remains essentially constant, while the volume increases with increasing strain (see Table \ref{PR}). This again demonstrates that the cut-off obtained under equilibrium conditions is transferable. Using the same amount of enclosed electrons seems logical, as for bulk material this value also remains constant.

\begin{table}[h!]
\centering

\caption{Poisson's ratio calculations of single-layer graphene. For all volumes $V(c)$ a cut-off $c$=0.0024 e$^-$/a$_0^3$ has been used. $N_Q=Q(c)/Q_{total}$ gives the ratio of enclosed electrons, $t$ the average equivalent slab thickness. $a,b$ are the relaxed in-plane orthorhombic lattice parameters in equilibrium using a supercell with 8 carbon atoms.}
\label{PR}

\begin{tabular}{l c c c c c }
\br
strain $\epsilon$ (\%) & $V(c)$ (a$_0^3$) & $a$ (a$_0$) & $b$ (a$_0$) & $t$ (\AA)& $N_Q$ (\%) \\
\mr
-2.0 & 454.53 &  7.836 & 9.269 & 3.31 & 99.65 \\
-1.0 & 458.07 & 7.916 & 9.249 & 3.31 & 99.64 \\
-0.5 & 459.96 & 7.956 & 9.241 & 3.31 & 99.64 \\
0.0 & 461.86 & 7.996 & 9.231 & 3.31 &  99.64 \\
+0.5 & 463.68 & 8.036 & 9.225 & 3.31 &  99.64 \\
+1.0 & 465.53 & 8.076 & 9.216 & 3.31 &  99.64 \\
+2.0 & 469.23 & 8.156 & 9.201 & 3.31 &  99.63 \\
\br
\end{tabular}
\end{table}


\end{document}